\documentclass[preprint,12pt]{elsarticle}
\usepackage{epsfig}
\usepackage{amsmath}
\usepackage{graphicx,color}
\usepackage{amsfonts}
\usepackage{amssymb}%
\usepackage{graphicx}

\begin{document}

\begin{frontmatter}

\title{Saturation and geometrical scaling -- \\
from small $x$ deep inelastic ep scattering \\ to high energy 
proton-proton \\ and heavy ion collisions \\
~~ \\
\footnotesize{Presented at the XXIII International Workshop on Deep Inelastic Scattering \\ and Related Subjects \\
		April 27 - May 1, 2015,
                Southern Methodist University,
		Dallas, Texas 75275}}

\author[jag]{Michal Praszalowicz}
\address[jag]{M. Smoluchowski Institute of Physics, Jagiellonian University, \\
ul. S. {\L}ojasiewicza 11, 30-348 Krak{\'o}w, Poland.}
        
\vspace{0.3cm}

\begin{abstract}
{Gluon distributions of colliding hadrons saturate as a result of the non-linear 
evolution equations of QCD. As a consequence  there exists the so called saturation 
momentum, which is related to the gluon density per unit rapidity per transverse area. 
When  saturation momentum is the only 
scale for physical processes, different observables exhibit 
geometrical scaling (GS). We  show a number of examples of GS and its violation in different reactions.}
\end{abstract}

\end{frontmatter}

\section{Deep inelastic scattering}

In this talk we summarize, following Ref.~\cite{Praszalowicz:2015hia}, our recent studies of
geometrical scaling in high energy collisions. Deep inelastic scattering (DIS) is well described
in terms of the dipole model (see {\em e.g.} \cite{GolecBiernat:1998js} and references therein):
\begin{equation}
\frac{F_{2}(x,Q^{2})}{Q^{2}}  =\frac{1}{4\pi^{3}}\int dr^{2}\left\{  \left\vert
\psi_{\rm T}(r,Q^{2})\right\vert ^{2}+\left\vert \psi_{\rm L}(r,Q^{2})\right\vert
^{2}\right\}  \,\sigma_{\rm{dp}}(r^{2})
\label{F2overQ2}
\end{equation}
where $\psi_{\rm T,L}$ are known functions that describe photon dissociation into a $q\bar{q}$
(dipole) pair. For massless quarks these functions have a property
\begin{equation}
\Phi_{\rm T,L}(u=rQ)=r^{2}\left\vert \psi_{\rm T,L}(r,Q^{2})\right\vert ^{2},%
\end{equation}
{\em i.e.} $\Phi_{\rm T,L}$ depend only on a combined variable $u$. Dipole-proton
cross-section $\sigma_{\rm{dp}}(r^{2})$ has to be modeled. If 
\begin{equation}
\sigma_{\rm{dp}}(r^{2})=\sigma_0 f(r^2 Q^2_{\rm s})
\label{sigGS}
\end{equation}
where $f$ is dimensionless function ($\sigma_0$ sets the dimension) of dipole size $r$ and
momentum scale $Q_{\rm s}$ then
\begin{equation}
{F_{2}(x,Q^{2})}/{Q^{2}}={\rm function}\left( {Q^2}/{Q_{\rm s}^2} \right).
\label{GSdef}
\end{equation}
Here $\tau = Q^2/Q^2_{\rm s}$ is called {\em scaling variable} and $Q_{\rm s}=Q_{\rm s}(x)$
denotes the {\rm saturation momentum}, which takes the following form
\begin{equation}
Q_{\rm{s}}^{2}=Q_{0}^{2}\left(  {x}/{x_{0}}\right)  ^{-\lambda
} \label{Qsat}%
\end{equation}
that follows from the traveling wave solutions \cite{MunPesch}
of the nonlinear QCD evolution equations \cite{jimwlk,BK}
In what follows only the value of exponent $\lambda$ will be of importance. 
Property (\ref{GSdef}) is called geometrical scaling \cite{Kwiecinski:2002ep}.

\begin{figure}[h]
\centering
\includegraphics[width=7.5cm]{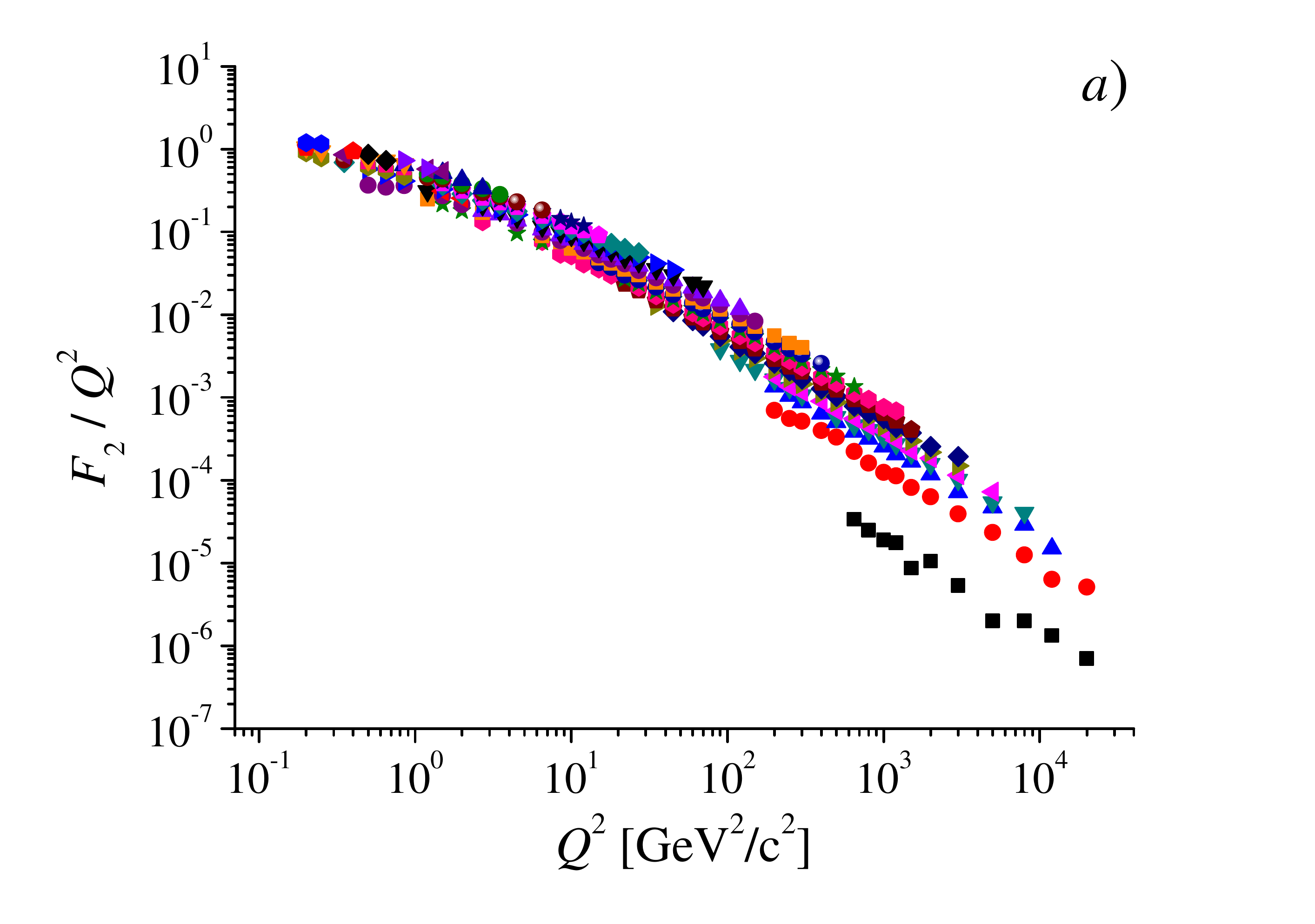}
\includegraphics[width=7.5cm]{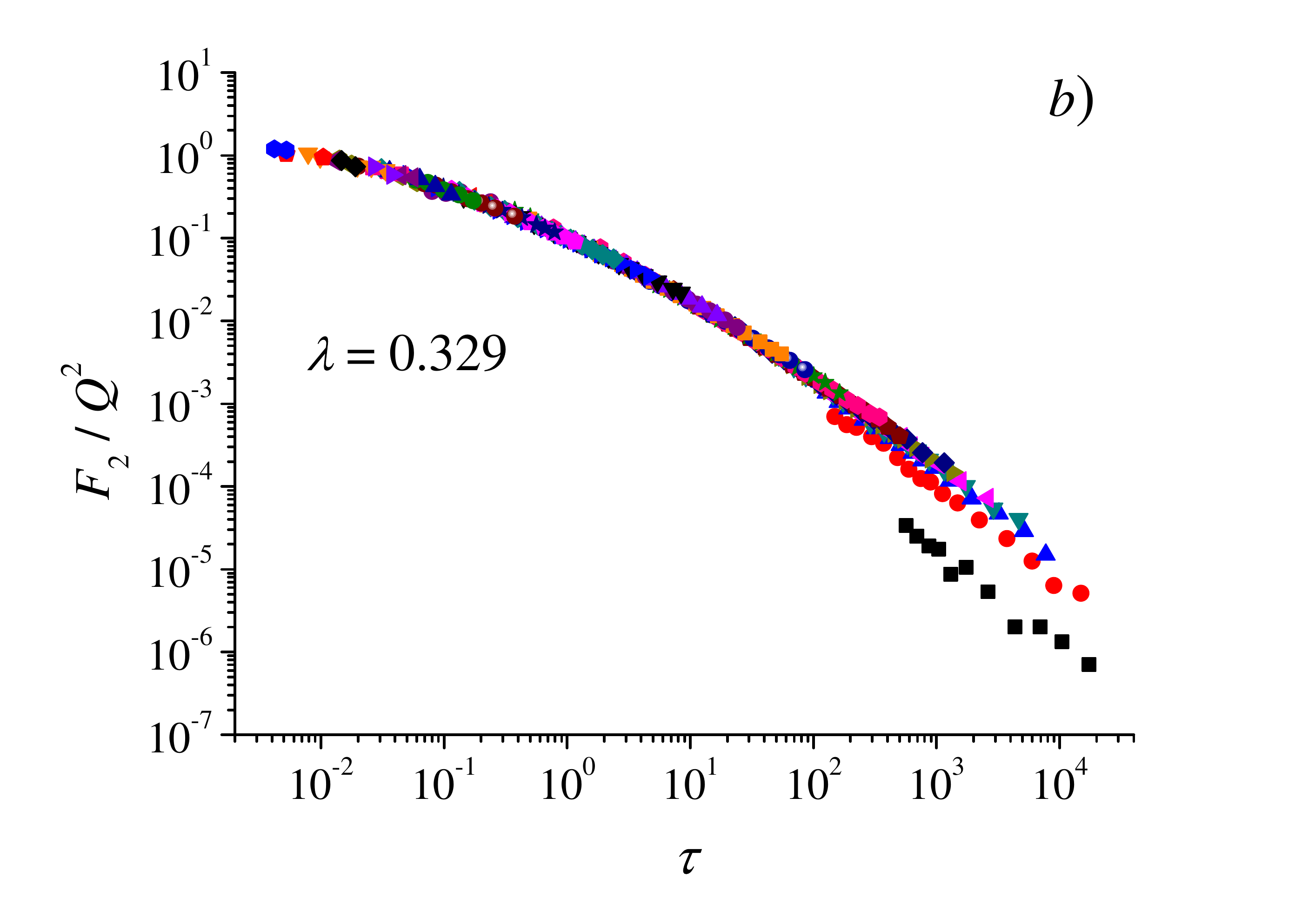}\caption{
Left: $\gamma^{\ast}p$ cross-sections $F_{2}/Q^{2}$ as
functions of $Q^{2}$ for fixed $x$. Different points correspond to different
Bjorken $x$'s. Right: the same but in function of scaling variable $\tau$ for
$\lambda=0.329$. Points in the right end of the plot correspond to large $x$'s
(due to kinematical correlation of the HERA phase space), and therefore show
explicitly violation of geometrical scaling. (Figure from Ref.~\cite{Praszalowicz:2012zh}.)}%
\label{figGSinx}%
\end{figure}

In Fig.~\ref{figGSinx} we show
that combined DIS data \cite{HERAcombined}
indeed exhibit GS. Exponent $\lambda$ has been extracted 
in a model-independent way in  Ref.~\cite{Praszalowicz:2012zh} and takes the
value of 0.329.

\section{Proton-proton scattering}

\begin{figure}[h!]
\centering
\includegraphics[width=6.8cm,angle=0]{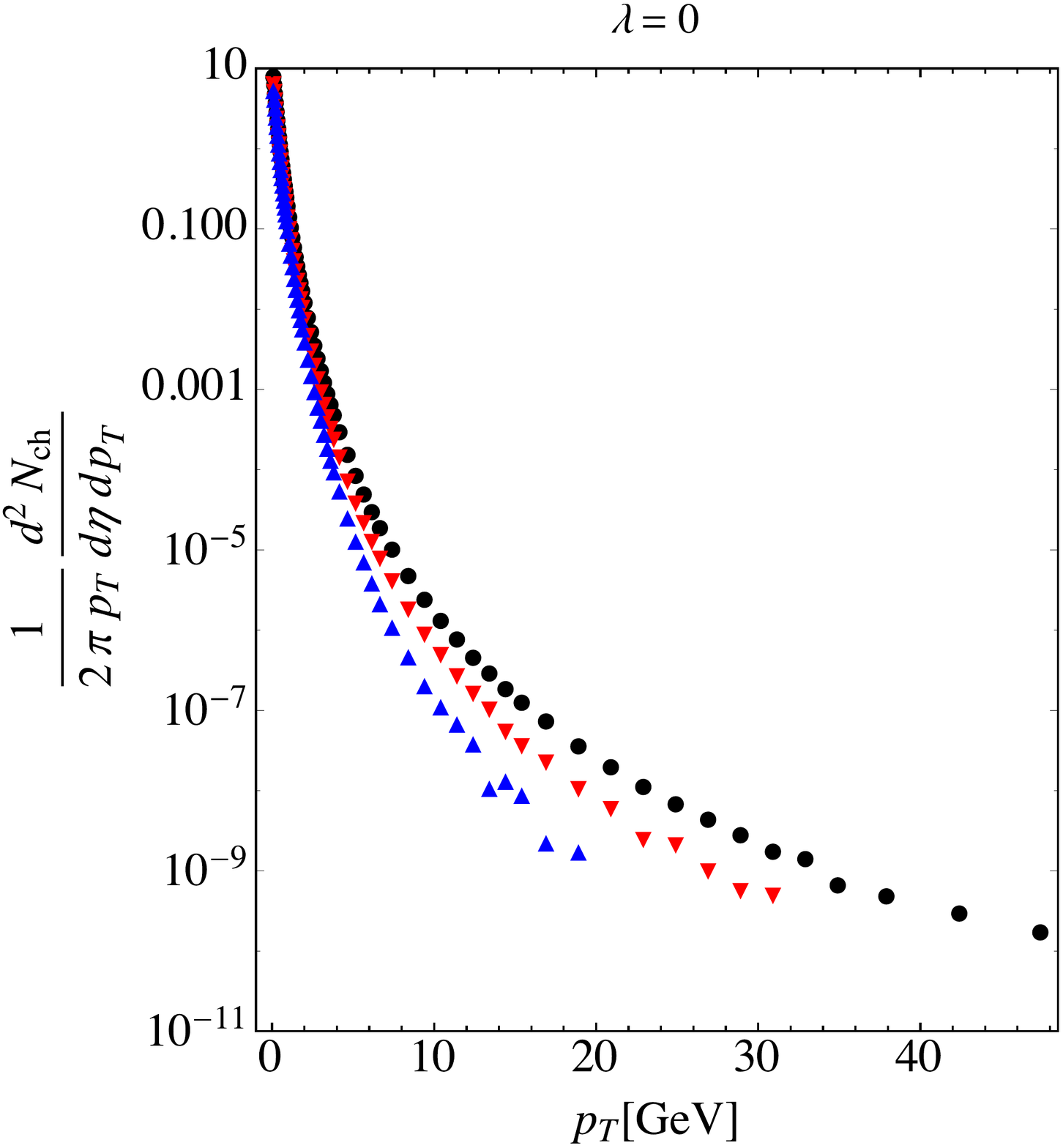}
\includegraphics[width=6.8cm,angle=0]{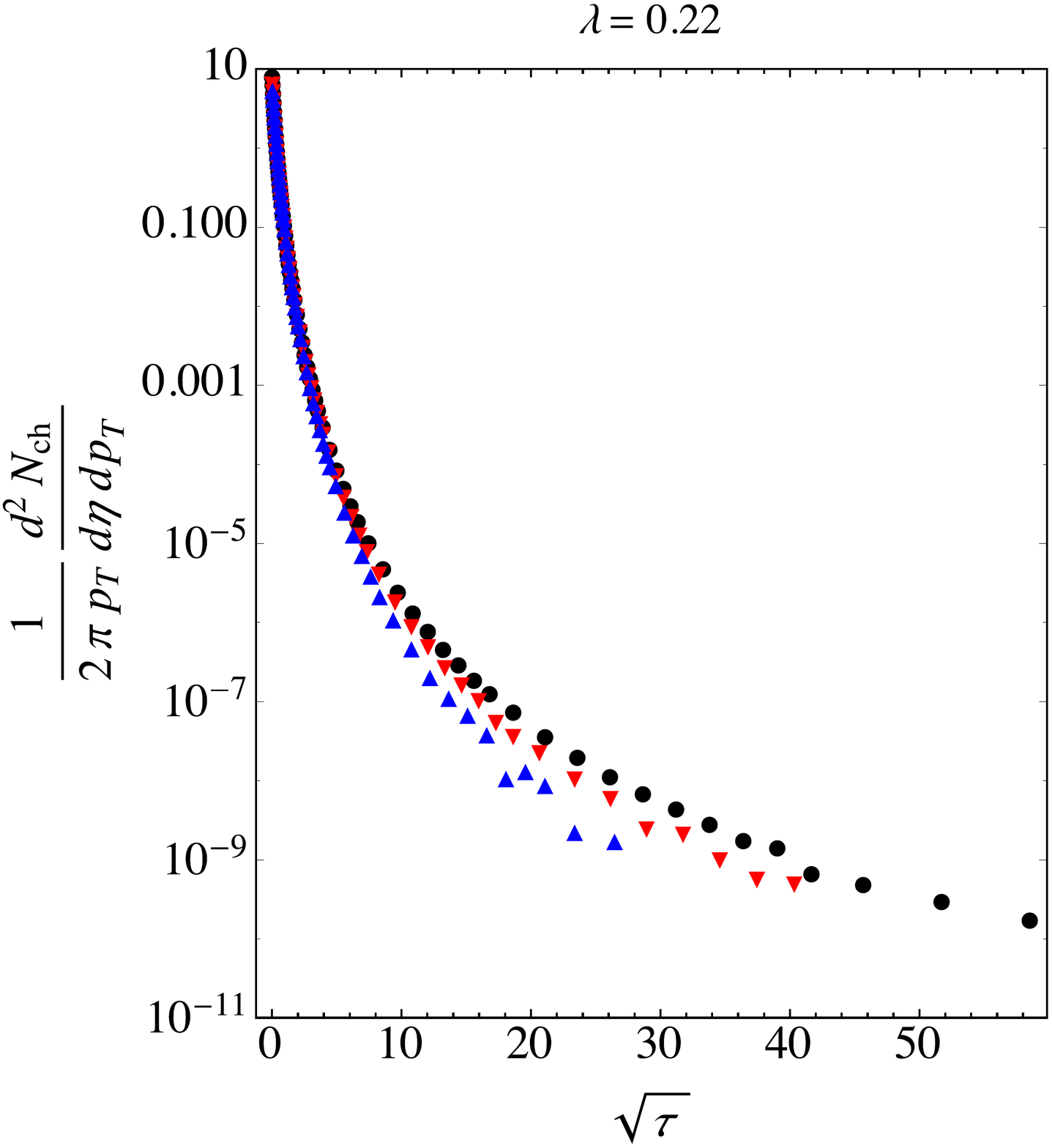} \caption{Data
for pp scattering from ALICE \cite{Abelev:2013ala} plotted in terms of
$p_{\mathrm{T}}$ and $\sqrt{\tau}$. Full (black) circles correspond to
$W=7$~TeV, down (red) triangles to 2.76~TeV and up (blue) triangles to
0.9~TeV.}%
\label{GSALICE}%
\end{figure}

GS in DIS follows from the scaling property (\ref{sigGS}) of the dipole cross-section, which
in turn is related to the unintegrated gluon distribution denoted in the following by 
$\varphi(k_{\rm T}^2,x)$. Inclusive gluon cross-section can be expressed in terms
of $\varphi$'s in the $k_{\rm T}$ factorization scheme \cite{Gribov:1981kg}:
\begin{equation}
\frac{d\sigma}{dyd^{2}p_{\rm T}}=\frac{3\pi}{2p_{\rm T}^{2}}%
{\displaystyle\int}
d^{2}\vec{k}_{{\rm T}}\,\alpha_{{\rm s}}(k_{{\rm T}}^{2})\varphi_{1}%
(x_{1},\vec{k}_{{\rm T}}^{2})\varphi_{2}(x_{2},(\vec{k}-\vec{p}\,)_{{\rm T}%
}^{2}).\label{sigma_def}%
\end{equation}
Here $\varphi_{1,2}$ are unintegrated gluon densities
and $x_{1,2}$ are gluon momenta
fractions needed to produce a gluon of transverse momentum $p_{{\rm T}}$ and rapidity
$y$:
\begin{equation}
x_{1,2}=e^{\pm y}p_{{\rm T}}/\sqrt{s}\label{x12}\, .%
\end{equation}

Note that unintegrated gluon densities have dimension of area. 
This is at best seen from the very simple parametrization
proposed by Kharzeev and Levin~\cite{Kharzeev:2001gp} or by 
Golec-Biernat and W{\"u}sthoff~\cite{GolecBiernat:1998js} in the context of DIS:%
\begin{equation}
\varphi_{\rm KL}(k_{{\rm T}}^{2})=S_{\bot}\left\{
\begin{array}
[c]{rrr}%
1 & {\rm for} & k_{{\rm T}}^{2}<Q_{{\rm s}}^{2}\\
Q_{{\rm s}}^{2}/k_{{\rm T}}^{2} & {\rm for} & k_{{\rm T}}^{2}<p_{{\rm T}%
}^{2}%
\end{array}
\right. \; \; \;{\rm or\ }\; \; \;\varphi_{\rm GBW}(k_{{\rm T}}^{2})=S_{\bot}\frac{3}{4\pi^{2}%
}\frac{k_{{\rm T}}^{2}}{Q_{{\rm s}}^{2}}\exp\left(  -k_{{\rm T}}%
^{2}/Q_{{\rm s}}^{2}\right)  .\label{glue}%
\end{equation}
Here $S_{\bot}$ is the transverse size given by inelastic cross-section (or
its part) for the minimum bias inclusive multiplicity or in the case of DIS
$S_{\bot}=\sigma_{0}$ is the dipole-proton cross-section for large dipoles.
Another feature of the unintegrated glue (\ref{glue}) is the fact that
$\varphi$ depends on the ratio $k_{{\rm T}}^{2}/Q_{{\rm s}}^{2}(x)$ rather
than on $k_{{\rm T}}^{2}$ and $x$ separately. 

An immediate consequence of (\ref{sigma_def}) is GS for the inelastic cross-section
in mid-rapidity ($y \sim 0$)
\begin{equation}
\frac{d\sigma}{dyd^{2}p_{{\rm T}}}=S_{\bot}^{2}\mathcal{F}(\tau
)\qquad{\rm or}\qquad\frac{1}{S_{\bot}}\frac{dN}{dyd^{2}p_{{\rm T}}%
}=\mathcal{F}(\tau)\label{sigma_2}%
\end{equation}
where $\tau=p_{{\rm T}}^{2}/Q_{{\rm s}}^{2}(x)$ is scaling variable. 
If 
\begin{equation}d\sigma =S_{\bot}dN \label{sigdN}
\end{equation}
 then second of Eqs.(\ref{sigma_2}) holds.
This implies that particle spectra $dN/dy$
at different energies should coincide if plotted in terms $\tau$.
In other words they should exhibit GS~\cite{McLerran:2010ex}
(if we neglect logarithmic violations of GS due to $\alpha_{\rm s}$
and assume parton-hadron duality \cite{Dokshitzer:1991eq}). That this
indeed happens is illustrated in Fig.~\ref{GSALICE}. The best quality
of GS is in this case achieved for $\lambda=0.22$, which is different
than $\lambda$ extracted from DIS.
In Ref.~\cite{FraPra}
we argue that this difference  is removed if
one assumes that the scaling observable is $d\sigma$ rather that $dN$, which
implies that the proportionality factor in Eq.~(\ref{sigdN}) is not 
energy independent $S_\bot$ but an inelastic cross-section $\sigma_{\rm in}(s)$.

Of course GS in pp is not perfect and extends only over the limited range
up to  $\sqrt{\tau} \sim 4$. Nevertheless it is still quite impressive, given the fact
that strictly speaking GS is a property of produced gluons. Physical particles
appear due to gluon fragmentation, they undergo final state interactions, and many
of them are in fact produced from resonance decays. All these effects seem to 
preserve GS.

As a consequence of Eq.~(\ref{sigma_2}) both integrated multiplicity $dN/dy$ and
 average transverse momentum $\langle p_{\rm T} \rangle$ grow as a power with 
energy \cite{McLerran:2010ex}.
This behavior is indeed seen in the data. Furthermore correlations of $\langle p_{\rm T} \rangle$
with $N_{\rm ch}$ are well described by GS supplemented  with model calculations within
Color Glass Condensate \cite{McLerran:2013una}.

\section{Heavy ion collisions}

\begin{figure}[h]
\centering
\includegraphics[width=6.6cm]{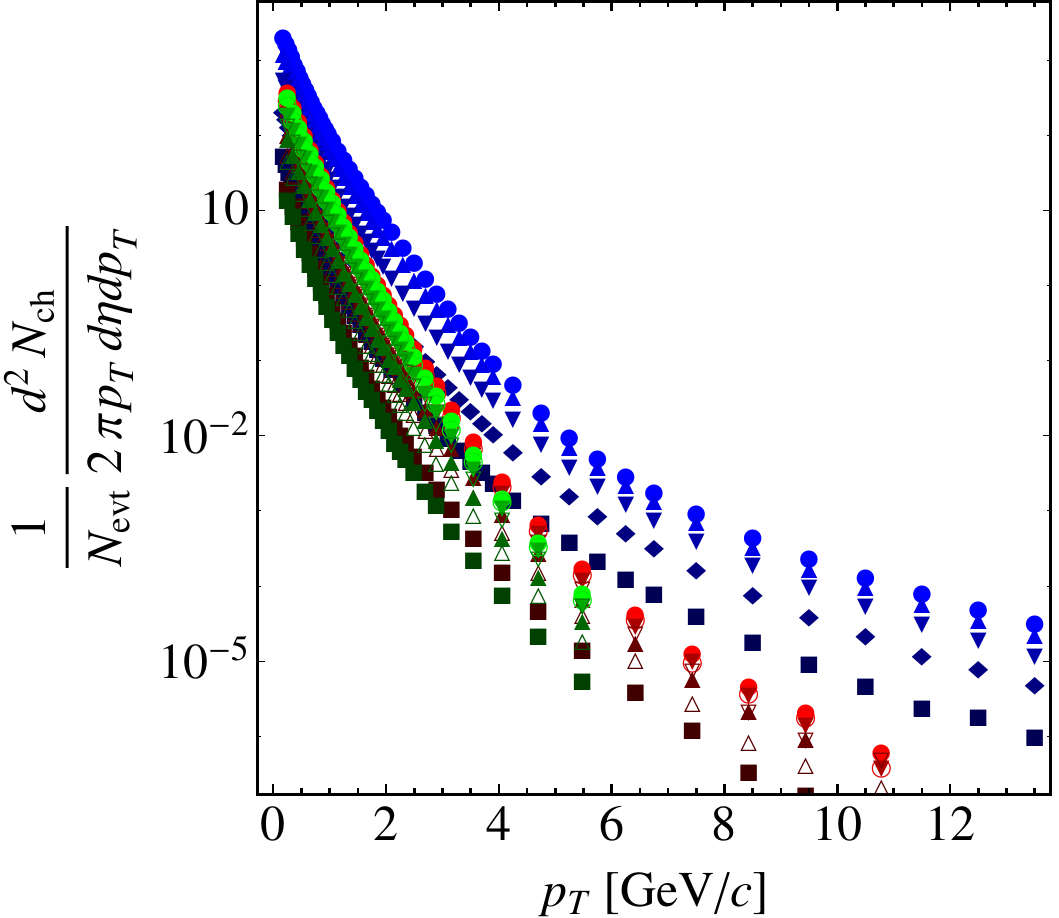}
\includegraphics[width=6.6cm]{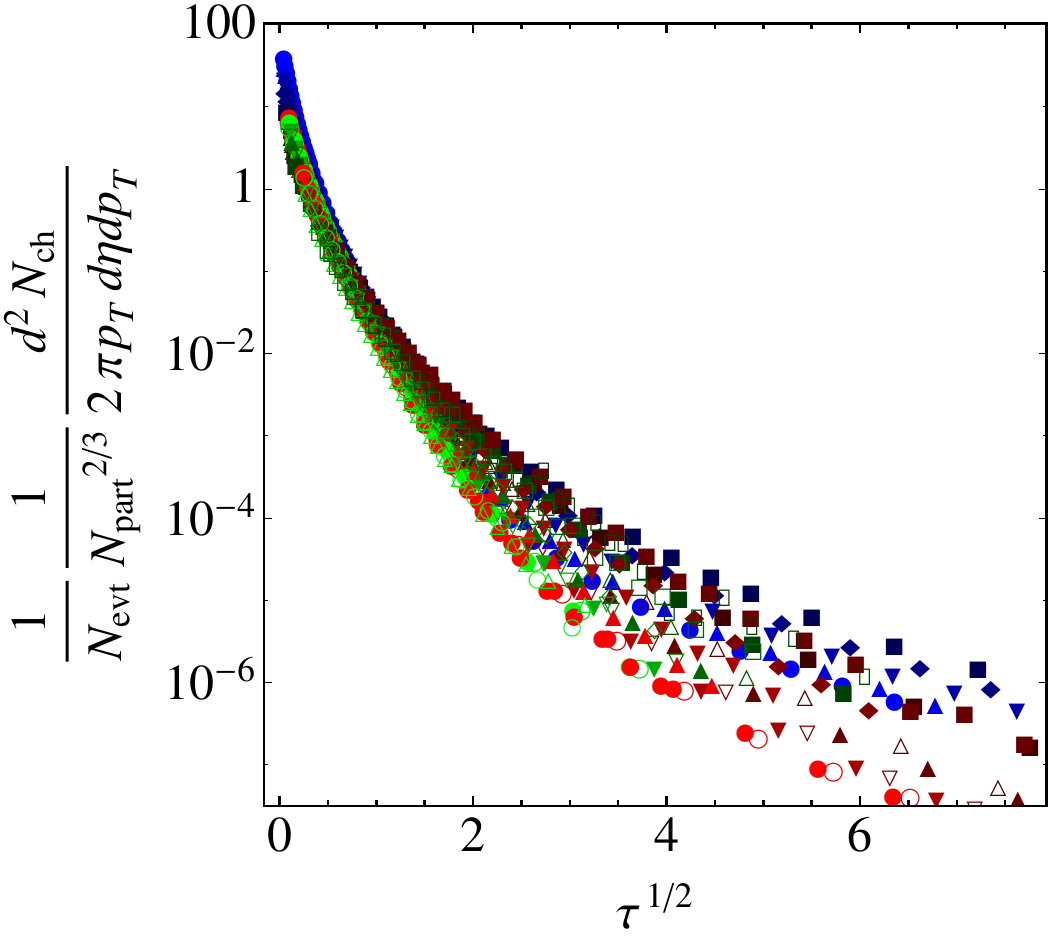}\caption{Illustration of
geometrical scaling in heavy ion collisions at different energies and
different centrality classes. Left panel shows charged particle distributions
from ALICE \cite{Abelev:2012hxa}, STAR \cite{Adams:2003kv,Adler:2002xw} and
PHENIX \cite{Adler:2003au,Adcox:2001jp} plotted as functions of $p_{\mathrm{T}%
}$. In the right panel the same distributions are scaled according to
Eq.~(\ref{multHI}). }%
\label{fig:all}%
\end{figure}

While GS in pp scattering -- as the property of the initial state -- might have 
come as a surprise, it would be even more so if GS were present in heavy
ion (HI) collisions. This is because strongly interacting matter undergoes
hydrodynamical evolution before it finally hadronizes. Nevertheless,
as we shall show below following Ref.~\cite{Praszalowicz:2015hia},
GS can be seen in particle spectra in HI collisions
both for hadrons \cite{Praszalowicz:2011rm}. 
GS holds also and for photons \cite{Klein-Bosing:2014uaa} 
that, however, probe the initial stage of the collision.

HI data are divided into centrality classes that select events within
certain range of impact parameter $b$. In this case both transverse area
$S_{\bot}$ and the saturation scale $Q_{\rm{s}}^{2}$ acquire additional
dependence on centrality that is characterized by an average number of
participants $N_{\rm{part}}$. We have
\cite{Klein-Bosing:2014uaa,Kharzeev:2004if}:%
\begin{equation}
S_{\bot}\sim N_{\rm{part}}^{2/3}\;\;\; \rm{and}\;\;\; Q_{\rm{s}}^{2}\sim
N_{\rm{part}}^{1/3}. \label{Npartscaling}%
\end{equation}
Therefore in HI collisions%
\begin{equation}
\frac{1}{N_{\rm{evt}}}\frac{dN_{\rm{ch}}}{N_{\rm{part}}^{2/3} \,d\eta
d^{2}p_{\rm{T}}}=\, \frac{1}{Q_{0}^{2}} \, F(\tau) \,\,\,\, \mathrm{where}
\,\,\,\, \tau=\frac{p_{\rm{T}}^{2}}{N_{\rm{part}}^{1/3} \, Q_{0}^{2}%
}\left(  \frac{p_{\rm{T}}}{W}\right)  ^{\lambda}. \label{multHI}%
\end{equation}
Note that by selecting certain centrality class we in fact select  an overlap $S_\bot$
between interacting ions and therefore one can safely use relation (\ref{sigdN}).

In Fig.~\ref{fig:all} we plot LHC and RHIC data in terms of $p_{\rm T}$ (left
panel) and $\sqrt{\tau}$ for $\lambda=0.3$ (right panel). One can see an
approximate scaling of, however, worse quality than in the pp case.

\section{Summary}

A wealth of data in ep and in hadronic collisions exhibits GS. In this note we have 
only mentioned some examples. The most important topics not included here are
extension of GS to the case of identified particles \cite{Praszalowicz:2013fsa}
and GS violation for $y \ne 0$ \cite{Praszalowicz:2013uu}.

GS may be
interpreted as a signature of saturation. However, one has to keep in mind that
it is a linear part of  QCD evolution equations that develops GS. Nonlinearities
serve as a "damping" that prevents scattering amplitudes  from growing over
the unitarity limit and -- at the same time --  making the entire solution to take asymptotically
the scaled form.

Many aspects of geometrical scaling require further studies. Firstly, new data 
from the LHC run II  (to come) have to be examined. On
theoretical side the universal shape ${\cal F}(\tau)$ has to be found and its
connection to the unintegrated gluon distribution has to be studied. That will
finally lead to probably the most difficult part, namely to the breaking of GS
in pp and in HI.

\section*{Acknowledgements} 
This research  has been financed in part by
the Polish NCN grant 2014/13/B/ST2/02486. 
The author wants to thank the organizers for a very successful meeting
and for financial support.

\end{document}